%% file: STmat.tex
\documentclass[reprint,aps,prl,superscriptaddress]{revtex4-1}
\pdfoutput=1

\usepackage{bm}
\usepackage{amsmath,amsthm}
\usepackage{mathtools,amssymb}
\usepackage{array,booktabs,longtable}
\usepackage{mathrsfs}
\usepackage{subfigure}
\usepackage{color}
\usepackage{tikz}
\usetikzlibrary{calc}

\newcommand{\ket}[1]{| #1\rangle}
\newcommand{\inp}[2]{\langle #1 |#2\rangle}
\newcommand{\inpp}[3]{\langle #1 |#2|#3\rangle}

\input{tikzDef}

\begin{document}

%\preprint{}

\title{Universal Wave Function Overlap and\\
Universal Topological Data from Generic Gapped Ground States}


%\author{}
%\thanks{}

\author{Heidar Moradi}
\affiliation{Perimeter Institute for Theoretical Physics, Waterloo, Ontario, N2L 2Y5 Canada}

\author{Xiao-Gang Wen}
\affiliation{Perimeter Institute for Theoretical Physics, Waterloo, Ontario, N2L 2Y5 Canada} 
\affiliation{Department of Physics, Massachusetts Institute of Technology, Cambridge, Massachusetts 02139, USA}

\date{\today}

\begin{abstract}
We propose a way -- universal wave function overlap -- to extract universal
topological data from generic ground states of gapped systems in any
dimensions. Those extracted topological data  should fully characterize the
topological orders with gapped or gapless boundary. For non-chiral topological orders in 2+1D, this universal topological data consist of two matrices,
$S$ and $T$, which generate a projective representation of $SL(2,\mathbb Z)$ on
the degenerate ground state Hilbert space on a torus.  For topological orders
with gapped boundary  in higher dimensions, this data constitutes a projective
representation of the mapping class group $\texttt{MCG}(M^d)$ of closed spatial
manifold $M^d$. For a set of simple models and perturbations in two dimensions, we show that these quantities are protected to all orders in perturbation theory.
\end{abstract}

%\keywords{}

\maketitle


Since the discovery the fractional quantum Hall effect
(FQHE)\cite{TSG8259,L8395} and theoretical study of chiral spin
liquids, \cite{KL8795,WWZcsp} it has been known that new kind of orders beyond
Landau symmetry breaking orders exist for gapped states of matter, called
topological order. \cite{Wtop,WNtop} Topological order can be thought of as the
set of universal properties of a gapped system, such as (a) the
topology-dependent \emph{ground state degeneracy} \cite{Wtop,WNtop} and (b) the
\emph{non-Abelian geometric phases $S$ and $T$} of the degenerate ground
states \cite{Wrig,KW9327,W1221}, which are \emph{robust against any local
perturbations} that can break any symmetries. \cite{WNtop} This is just like
superfluid order which can be thought of as the set of universal properties:
zero-viscosity and quantized vorticity, that are robust against any local
perturbations that preserve the $U(1)$ symmetry.  It was proposed that the
non-abelian geometric phases of the degenerate ground states on the torus
classify 2+1D topological orders. \cite{Wrig}

Interestingly, it turns out that non-trivial topological order is related to
long-range quantum entanglement of the ground state \cite{CGW1038}. These
long-range patterns of entanglement are responsible for the interesting physics,
such as quasiparticle excitations with exotic statistics, completely robust
edge states, as well as the universal ground state degeneracy and non-Abelian
geometric phases mentioned above.

Our current understanding is that topological order 
%\footnote{\textcolor{red}{Mention examples of stable gapped phases, which are
%beyond Atiyah's TQFT framework such as \cite{Yoshida:2013sqa}?}} 
in 2+1 dimensions is characterized by a unitary modular tensor category (UMTC)
which encode particle statistics and gives rise to representations of the Braid
group, \cite{Wang10} and the chiral central charge $c_-$ which encode
information about chiral gapless edge states. \cite{Wedgerev,Wtoprev} 

While the algebraic theory of 2+1D topological order is largely understood, it
is natural to ask whether it is possible to extract topological data from a
generic non-fixed point ground state.  One such proposal has been through using
the non-Abelin geometric phase $S$ and
$T$. \cite{Wrig,KW9327,W1221,ZGT1251,CV1223,ZMP1233,TZQ1251} Another is using
the entanglement entropy \cite{Kitaev:2005dm,Levin:2006zz} which has the
generic form in 2+1 dimensions $S = \alpha L - \gamma + \mathcal O(\frac 1L)$,
where $\gamma$ is the topological entanglement entropy (TEE). It turns out that
$\gamma = \log\mathcal D$, where $\mathcal D$ is the total quantum dimension
and thus a topological property of the gapped phase. A generalization of TEE to
higher dimensions were proposed in \cite{2011PhRvB..84s5120G}.
%A TQFT in the continuum limit, such as a Chern-Simons theory, captures the
%pure universal piece and therefore $S = -\gamma$, while for a generic
%dynamical theory non-universal contributions are non-zero.  The topological
%entanglement entropy, however, only captures a small piece of the data needed
%to characterize the topological order of a ground state. The question is, can
%we from a non-fixed point ground state extract more data that could
%characterize the underlying TQFT more fully?

Here, we would like to  propose a simple way to extract data from non-fixed point
ground states, that could fully characterize the underlying TQFT.  We conjecture
that for a system on a $d$-dimensional manifold $M^d$ of volume $V$ with the
set of degenerate ground states $\{\ket{\psi_\alpha}\}_{\alpha=1}^N$, the overlaps
of the degenerate ground states have the following form \cite{HW1339,KW13}
\begin{equation}\label{eq:overlap}
	\inpp{\psi_\alpha}{\hat {\mathcal O}_A}{\psi_\beta} = \:e^{-\alpha V + o(1/V)}\: M^A_{\alpha,\beta},
\end{equation}
where $\hat {\mathcal O}_A$, labeled by index $A$, are transformations of the
wave functions induced by the automorphism transformations of the space $M^d
\to M^d$, $\alpha$ is a non-universal constant, and $M^A$ is an
\emph{universal} unitary matrix (upto an overall $U(1)$ phase).  $M^A$ form a
projective representation of the automorphism group of the space $M^d$ --
$\texttt{AMG}(M^d)$, which is robust against any perturbations.  We propose
that \emph{such projective representations for different space topologies are
the universal topological data that fully characerize topological orders with
finite ground state degeneracy.} The disconnected components of  the
automorphism group is the mapping class group: $\texttt{MCG}(M^d) \equiv
\pi_0[\texttt{AMG}(M^d)]$.  We propose that \emph{ projective representations
of  the mapping class group for different space topologies are the universal
topological data that fully characerize topological orders with gapped
boundary.} (For a more general and a more detailed discussion, see Ref.
\cite{KW13}.) 

If the space is a 2D torus $T^2$, then the torus mapping class group
$\texttt{MCG}(T^2)=SL(2,\mathbb Z)$ is generated by a $90^\circ$ rotation $\hat
S$ and a Dehn twist $\hat T$. The corresponding $M^A$ are the unitary matrices
$S,\ T$ which generate a projective representation of $SL(2,\mathbb Z)$.
Compared to the proposal in Ref. \cite{Wrig,KW9327,W1221}, here we do not need
to calculate the geometric phase for a family of ground states and only have to
consider a much simpler calculation -- a particular overlap (with the cost of a
non-universal contribution with volume scaling). We will calculate this for the
simple example of $\mathbb Z_N$ topological state studied in Refs.
\cite{RS9173,Wsrvb,BB0781,SchulzetAlZn,YW1207} and investigate the universality of this
under perturbations such as adding string tension.

We note that a UMTC that describe the statistics of the exications in $2+1$D, can also
gives rise to a projective representation of $SL(2,\mathbb Z)$.  We propose
that the universal wave function overlap eqn. (\ref{eq:overlap}) computes this
projective representation.  The representation is generated by two elements $S$
and $T$ satisfying the relations
\begin{equation}\label{eq:SL2ZGeneratorRelations}
	(ST)^3 = e^{\frac{2\pi i}8c_-}\:C,\quad S^2=C,
\end{equation}
where $C$ is a so-called charge conjugation matrix and satisfy $C^2=1$.
Furthermore we have that $\frac 1{\mathcal D}\sum_a d_a^2\theta_a =
e^{\frac{2\pi i}8c_-}$, where $d_a$ and $\theta_a$ are the quantum dimension
and topological spin of quasiparticle $a$, respectively. This shows that the
UMTC, or particle statistics, fixes the chiral central charge$\mod 8$.
\footnote{The ambiguity of $c_-$ can be understood by the existence of the
so-called $E_8$ state, which can be realized by a Chern-Simons theory where the
$K$ matrix is the Cartan matrix of $E_8$. This theory has only trivial bulk
excitations since $\det K=1$ but boundary theory given by the affine Lie
algebra $(\hat E_8)_1$, which has $c_- =8$. Thus there is always the ambiguity
of adding a $E_8$ state without changing the bulk excitations, but shifting the
chiral central charge by $8$.  The chiral central charge is related to
perturbative gravitational anomalies on the edge, which signals lack of energy
conservation, or a gravitational parity anomaly from the bulk perspective.
Physically, this corresponds to a thermal Hall effect by the Callan-Harvey
inflow mechanism \cite{Callan:1984sa} and is a consequence of the decent
relations of anomalies in different dimensions. Note that in the case $c_- =
0$, the edge states are not chiral but they can however still be completely
robust. \cite{L1309} This is related to global gravitational anomalies, ie
modular anomaly on the edge.} This constitutes a projective representation of
$SL(2,\mathbb Z)$ on the groundstate subspace on a torus, which encode how the
groundstates transform under large automorphisms $\texttt{MCG}(T^2)$.  We
believe that our higher dimensional universal quantities \eqref{eq:overlap}
also encode information about the topological order in the ground state. 

%More importantly for us, it was
%believed that from the modular $S$ and $T$ matrices, 
%in a particular basis, to be called the  basis, 
%one can reconstruct the UMTC.\cite{Wang10,Wrig}

%In particular, the fusion rules can be reconstructed using the Verlinde
%formula \[\mathcal N^c_{ab} = \sum_x\frac{S_{ax}S_{bx}S^*_{cx}}{S_{1x}},\] and
%then solving the pentagon and hexagon equations will give all the information
%of the UMTC.

\emph{Construction of degenerate set of ground states from local tensor
networks}:
Since topological order exist even on topologically trivial manifolds, all its
properties should be available from a local wave function. But we need to
sharpen what we mean by a local wave functions, since wave functions typically
depend on global data such as boundary conditions. Amazingly, there exist a
surprisingly simple local representation of globally entangled states using
tensor network language. In particular, a tensor network state (TNS) known as PEPS, is
given by associating a tensor $T^{[i]}_{\sigma_i}(\alpha\beta\gamma\dots)$ to
each site $i$, where $\sigma_i$ is a physical index associated to the local
Hilbert space, and $\alpha,\beta,\gamma$ are inner indices and connect to each
other to form a graph. Using this local representation, the wave function is
then given by
\begin{equation}
	\ket{\psi} = \sum_{\{\sigma_i\}}\text{tTr}\left(T^{[1]}_{\sigma_1}T^{[2]}_{\sigma_2}\dots\right)\ket{\sigma_1,\sigma_2,\dots},
\end{equation}
where $\text{tTr}(\dots)$ contracts the tensor indices in the tensor product network.  By choosing the dimension of the inner indices large enough, one can approximate any state arbitrarily well. This particular representation is especially interesting for the study of gapped states since it automatically satisfies the area law, a property gapped ground states are known to have  \cite{Hastings2007,Masanes:2009tg}. Thus one can think of a TNS as a clever way of parametrizing the interesting sub-manifold of the Hilbert space, where ground states of local gapped Hamiltonians live.

A local tensor representation of a wave function however is not enough, it must be equipped with a gauge structure \cite{2010PhRvB..82p5119C}. Surprisingly, local variations of a tensor do not always correspond to local perturbations of the Hamiltonian and can change the global topological order. In order to approximate the ground state of a Hamiltonian with topological order with gauge group $G$, it is important to search within the set of variational tensors with symmetry $G$. Arbitrarily small $G$ breaking variations, might lead to tensor networks which can approximate local properties of a system well but give wrong predictions about the global properties.

In \cite{2010arXiv1001.4517S} a few concepts were introduced to characterize the symmetry structure of a TNS. In particular $d_{\text{space}}$-IGG, which is the group of intrinsically $d_{\text{space}}$-dimensional gauge transformations on the inner indices that leave the tensors invariant. It was in particular shown that in the case of the two-dimensional $\mathbb Z_2$ topological state we have $2$-IGG $= \mathbb Z_2$. Furthermore it was shown that $2$-IGG contains information about string operators and can be used to construct the full set of degenerate ground states on the torus from a local tensor representation.
\footnote{A related concept for a finite group $G$, is $G$-injective PEPS \cite{SchuchEtal}. A $G$-injective tensor is a tensor which is invariant under a $G$-action on all inner indices simultaneously, together with the property that one can achieve any action on the virtual indices by acting on the physical indices. It was shown that these tensors are ground states of a parent Hamiltonian and have the topological entanglement entropy $\gamma = \log |G|$. This class of PEPS describe the universality class of quantum double models $D(G)$. Recently this concepts was generalized to $(G,\omega)$-injective PEPS \cite{Buerschaper:2013nga}, where the action of $G$ on the tensors are twisted by a 3-cocycle of $\omega$ of $G$. It was shown that these PEPS describes topological order in the university class of Dijkgraaf-Witten TQFT's \cite{Dijkgraaf:1989pz} and only depend on the cohomology class $[\omega]\in H^3(G,U(1))$ of $\omega$.}

%The point of this discussion is the following. Imagine one is interested in probing whether a 2D local gapped Hamiltonian has topological order described by gauge group $G$. It is then important to search within the set of variational tensors with $2$-IGG$ = G$ and keep this symmetry intact during all variations and along renormalization group flows. This is a necessary symmetry condition to probe the topological order \cite{2010PhRvB..82p5119C}. If the ground state of the system has topological order with gauge group $G$, how can we find out which kind of topological order it has? For example, for $G=\mathbb Z_2$ there are two theories with topological order, the universality classes corresponding to the $\mathbb Z_2$ toric code and the double semion model.

%This can be decided if one knows the corresponding modular $S$ and $T$ matrices. First one can construct the set of ground states on a torus from the local tensor, by exploiting its gauge structure as in \cite{2010arXiv1001.4517S}. Now one can extract the $S$ and $T$ matrices by considering the overlaps \eqref{eq:overlap}. This will also fix the chiral central charge $c_-$, modulo $8$.

Thus the local data we need is = local tensor + gauge structure. From this gauge structure we can twist the tensor to get the full set of ground states on a torus \cite{2010arXiv1001.4517S,SchuchEtal,Buerschaper:2013nga}. We shall call the natural basis we get from such a procedure for twist basis.

\begin{figure}[tb]
\centering
\subfigure[\ \label{fig:Lattice}]{
\includegraphics[width=.10\textwidth]{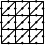}}\qquad
\subfigure[\ \label{fig:ZnTensorNetwork} ]{
\includegraphics[width=.11\textwidth]{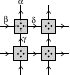}}\qquad
\subfigure[\ \label{fig:ZnTensorSymmetry} ]{
\includegraphics[width=.18\textwidth]{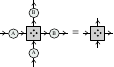}}
\caption{(a) Lattice under consideration, with the spins living on the links. (b) Tensor network for $\mathbb Z_N$ gauge theory. The lattice is chosen with the orientation shown. The tensors live on the lattice sites and the dots represent the physics indices. (c) Symmetry of the $\mathbb Z_N$ tensor.}
\end{figure}

We will in the following consider the $\mathbb Z_N$ topological state.  We can construct a local tensor for this state in the following way. Let the physical spins live on the links of the lattice, and give each link an orientation as in figure \ref{fig:ZnTensorNetwork}. Put a tensor $T^{(\sigma_1\sigma_2\sigma_3\sigma_4)}_{\alpha\beta\gamma\delta}$ on each site and require that
\begin{equation}\label{eq:ZnTensor}
	T^{(\alpha\beta\gamma\delta)}_{\alpha\beta\gamma\delta} = 1  \quad\text{if } \beta+\gamma-\alpha-\delta = 0\mod N,
\end{equation}
otherwise $T^{(\sigma_1\sigma_2\sigma_3\sigma_4)}_{\alpha\beta\gamma\delta}=0$. This tensor has a $\mathbb Z_N$ symmetry given by the tensors (see figure \ref{fig:ZnTensorSymmetry})
\[
\raisebox{-.2\height}{\includegraphics[width=0.10\textwidth]{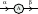}}
= \delta_{\alpha\beta}e^{\frac{2\pi i}N\alpha},\quad\raisebox{-.2\height}{\includegraphics[width=0.10\textwidth]{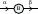}}
= \delta_{\alpha\beta}e^{\frac{-2\pi i}N\alpha}.
\]

%\emph{$S$ and $T$ from ground states -- a $\mathbb Z_N$ example}:
\emph{$\mathbb Z_N$ Topological Order}:
Equipped with the ground states of from local tensors, one can calculate the overlap \eqref{eq:overlap} to extract the universal topological properties.

As a simple example, let us calculate the overlap \eqref{eq:overlap} for the
case of $\mathbb Z_N$ topological state on the lattice in figure
\ref{fig:Lattice}. For this simple example we will not use tensor product
representation for simplicity, since it can be calculated directly. See \cite{HeMoradiWen} for calculation of \eqref{eq:overlap} using tensor network and gauge structure.

Let there be a local Hilbert space $\mathcal H_a\approx\mathbb C[\mathbb Z_N]\approx\mathbb C^N$ associated to each link $a\in\Omega$ with basis $\{\ket{\sigma_a}\}_{\sigma_a=0}^{N-1}$.

We will represent a spin configuration $\ket{\sigma_{a_1}\sigma_{a_2}\dots}$ using a string picture, where the state on link $a\in\Omega$ is represented by an oriented string of type $\sigma_a\in\mathbb Z_N$ with a chosen orientation, and $\ket 0$ corresponds to no string.
There is a natural isomorphism $\mathcal H_a\overset{\sim}{\rightarrow}\mathcal H_{a^\star}$ for link $a$ and its reversed orientation $a^\star$ by $\ket{\sigma_a}\mapsto\ket{\sigma_{a^\star}}=\ket{-\sigma_a}$.

The ground state Hilbert space of the $\mathbb Z_N$ topological order consists of an equal superposition of all closed-string configurations that satisfy the $\mathbb Z_N$ fusion rules.

The string-net ground state Hilbert space on $T^2$ can be algebraically constructed in the following way. Let $\Lambda^\star_\bigtriangleup$ denote the set of triangular plaquettes and for each $p\in\Lambda^\star_\bigtriangleup$ define the string operator $B^\bigtriangleup_p$ which act on the links bounding $p$, with clockwise orientation, by $\ket\sigma\mapsto\ket{\sigma+1\text{ mod N}}$. The set of all contractable closed loop configurations can be thought of as the freely generated group $G_{\text{free}} = \left<\{B^\bigtriangleup_p\}_{p\in\Lambda_\bigtriangleup^\star}\right>$, modulo the relations $\left(B^\bigtriangleup_p\right)^N \sim 1$, $\prod_{p\in\Lambda_\bigtriangleup^\star}B^\bigtriangleup_p \sim 1$ and $B^\bigtriangleup_pB^\bigtriangleup_q \sim B^\bigtriangleup_qB^\bigtriangleup_p$,
denoted as $G^{00}_\bigtriangleup = G_{\text{free}}/\sim$. Similarly we let the subgroup $G^{00}_\square\subset G^{00}_\bigtriangleup$ correspond to closed loop configurations on the square lattice links. For the ground states on the torus, we need to introduce two new operators $W_x$ and $W_y$, corresponding to non-contractable loops along the two cycles of $T^2$. These satisfy $(W_i)^N = 1$, $i=x,y$. With these, we can construct the group $G^{\alpha\beta}_{\bigtriangleup}$, corresponding to closed string configurations with $(\alpha,\beta)$ windings around the cycle $(x,y)$, modulo $N$. Similarly, let $G_{\bigtriangleup}$ be the group of all possible closed string configurations on the torus. These states are orthonormal $\inp {g_{\alpha\beta}}{\bar g_{\bar\alpha\bar\beta}} = \delta_{g_{\alpha\beta},\bar g_{\bar\alpha\bar\beta}}$.

The $N^2$-dimensional ground state Hilbert space is then spanned by the following vectors $\ket{\alpha,\beta} = |G^{\alpha\beta}_\bigtriangleup|^{-1/2}\sum_{g_{\alpha\beta}\in G^{\alpha\beta}_\bigtriangleup}\ket{ g_{\alpha\beta}}$,
where $\alpha,\beta = 0, \dots, N-1$.
The construction can trivially be extended to higher-genus surfaces.

This is the string-net basis for the $\mathbb Z_N$ gauge theory. The ground states in the twist basis corresponding to the tensor \eqref{eq:ZnTensor}, are just the eigenbasis the operators $W_x$ and $W_y$. These are given by
\begin{equation}
	\ket{\psi_{ab}} = \frac 1{\sqrt{|G_\bigtriangleup|}}\sum_{g\in G_{\bigtriangleup}}\gamma^{a\omega_x(g) + b\omega_y(g)}\ket g,
\end{equation}
where $\gamma = e^{-\frac{2\pi i}N}$ and $\omega_i$ count how many times the string configuration $g$ wraps around the $i$'th cycle. Note that $W_x\ket{\psi_{ab}} = e^{\frac{2\pi i}N a}\ket{\psi_{ab}}$ and $W_y\ket{\psi_{ab}} = e^{\frac{2\pi i}N b}\ket{\psi_{ab}}$.
For later use, note that $|G^{\alpha\beta}_{\bigtriangleup}| =N^{|\Lambda^\star_\bigtriangleup|-1}=N^{2L^2-1}$, $|G^{\alpha\beta}_{\square}| =N^{|\Lambda^\star_\square|-1}=N^{L^2-1}$, $|G_{\bigtriangleup}| =N^2 |G^{\alpha\beta}_{\bigtriangleup}|$ and $|G_{\square}| =N^2 |G^{\alpha\beta}_{\square}|$.

\begin{figure}[tb]
\centering
\includegraphics[width=.48\textwidth]{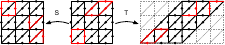}
\caption{Definition of $S$ and $T$ transformations. The $S$ transformation corresponds to rotating configurations $90$ degrees, while $T$ corresponds to a shear transformation. Note that this transformation does not leave the space of closed loop configurations invariant.}\label{fig:SandT}
\end{figure}

\emph{Modular $S$ and T-matrix from the ground state}:
We can now define two non-local operators on our Hilbert space $\hat{\mathcal
O}_S, \hat{\mathcal O}_T:\mathcal H\rightarrow\mathcal H$ as in figure
\ref{fig:SandT}, mimicking the generators of the torus mapping class group in
the continuum. Here $\hat{\mathcal O}_S$ maps any spin configuration, to the
$90$ degree rotated configuration. $\hat{\mathcal O}_T$ corresponds to shear
transformation and is defined as in figure \ref{fig:SandT}.  It is clear that
since we are on the lattice, these operators will not preserve the subspace of
closed string configurations.

We can easily calculate the matrix elements of $\hat{\mathcal O}_T$ and $\hat{\mathcal O}_S$ between ground states. In both cases, only $|G_\square|$ configurations have a non-zero overlap with the un-deformed ground state. For the $S$ transformation we find the overlap
\begin{align*}
	&\inpp{\psi_{ab}}{\hat {\mathcal O}_S}{\psi_{\bar a\bar b}}
		= \delta_{a,\bar b}\delta_{b,-\bar a}\frac{|G_\square|}{|G_{\bigtriangleup}|} = S_{ab,\bar a\bar b}  e^{-\log(N)L^2},
\end{align*}
where we have defined the modular $S$ matrix
$S_{ab,\bar a\bar b} = \delta_{a,\bar b}\delta_{b,-\bar a}$.
Similarly we have $\inpp{\psi_{ab}}{\hat {\mathcal O}_T}{\psi_{\bar a\bar b}}									  = T_{ab,\bar a\bar b}  e^{-\log(N)L^2},$
where the modular $T$ matrix is given by
$T_{ab,\bar a\bar b} = \delta_{a+b,\bar a}\delta_{b,\bar b}$. 
One can readily check that these satisfy eq. \eqref{eq:SL2ZGeneratorRelations} with $c_-=0\mod 8$ and $C_{ab,\bar a\bar b} = \delta_{a,-\bar a}\delta_{-b,\bar b}$. Thus this forms a projective representation of the modular group $SL(2,\mathbb Z)$.

In order to use Verlinde's formula and generate the relevant UMTC, we need to put the modular matrices in the quasi-particle basis \footnote{In general we do not have a gauge theory and need another way to find the modular matrices in the right basis. One way is to find the basis which diagonalize $T$ and the $S$ matrix satisfy the requirements
\begin{align*}
	&1.\;\; \mathcal N^c_{ab} = \sum_x\frac{S_{ax}S_{bx}S^*_{cx}}{S_{1x}}\;\text{ is a positive integer},\\
	&2.\;\; S_{ab} = S_{ba},\\
	&3.\;\; S_{1,a}>0.
\end{align*}
In \cite{2013arXiv1303.0829L} it was shown for several examples, that this
basis is unique and lead to the right form of $S$ and $T$.}. This is done as
follows, for the $\mathbb Z_N$ theory there are non-contractable magnetic
operators on the dual lattice satisfying $(\Gamma_i)^N = 1$, and with the
commutation relations $W_x\Gamma_y = e^{-\frac{2\pi i}N}\Gamma_y W_x$ and
$W_y\Gamma_x = e^{-\frac{2\pi i}N}\Gamma_x W_y$. The basis we are after
corresponds to having a well-defined magnetic and electric flux through one
direction of the torus. In the eigenbasis of $W_y$ and $\Gamma_y$,
$\ket{\phi_{mn}}$, we find
\begin{align*}
	S_{mn,\bar m\bar n} = \frac 1Ne^{-\frac{2\pi i}N(m\bar n+n\bar m)},\quad T_{mn,\bar m\bar n} = \delta_{m,\bar m}\delta_{n,\bar n} e^{\frac{2\pi i}Nmn}.
\end{align*}
These are the well-known modular matrices for the $\mathbb Z_N$ model.

\begin{figure}[tb]
\centering
\includegraphics[scale=0.40]{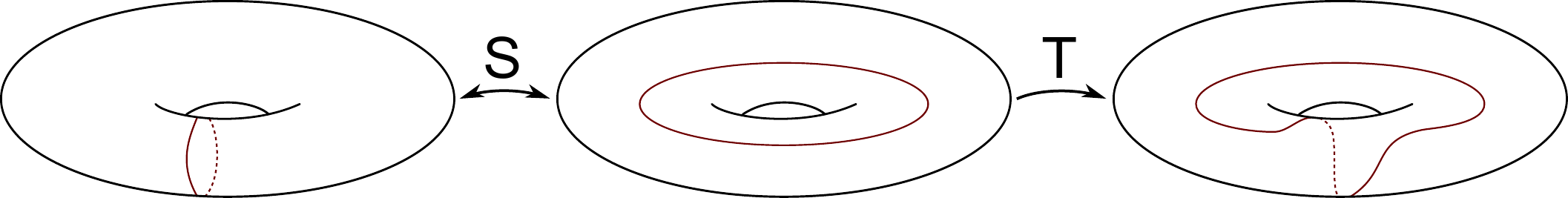}
\caption{In the string-net basis, a modular $S$ transformation flips the topological sectors $(\alpha,\beta)\rightarrow (\beta,-\alpha \mod N)$, while a $T$ transformation has the effect $(\alpha,\beta)\rightarrow (\alpha,\alpha+\beta\text{ mod } N)$.}
\end{figure}

\emph{Perturbed $\mathbb Z_N$ model}:
We will now consider a local perturbation to the $\mathbb Z_N$ topological state. One interesting perturbation is to add a magnetic field of the form $\frac J2\sum_{a\in\Omega}(Z_a+Z_a^\dagger)$, where $Z_a$ is a local operator defined as $Z_a\ket{\sigma_a}=e^{\frac{2\pi i}N\sigma_a}\ket{\sigma_a}$ \footnote{See \cite{SchulzetAlZn} for analysis of models of this type.}. This perturbation breaks the exact solvability of the model, but essentially corresponds to introducing string tension to each closed string configuration. This can be implemented by local deformation of the ground states of the form
\begin{equation*}
	\ket{\psi_{ab}}_{\mathcal A} = \frac 1{\sqrt{|G_\bigtriangleup|}}\sum_{g\in G_\bigtriangleup}\mathcal A^{-\mathcal L(g)/2}\gamma^{a\omega_x(g)+b\omega_y(g)}\ket g,
\end{equation*}
where $\mathcal A$ is a variational parameter. %$\mathcal L(g)$ is essentially the total string length of the closed string configuration $g$, but any string of type $\sigma$ is weighted by $\frac 12[1-\cos(\frac{2\pi}N\sigma)]$.
Furthermore $\mathcal L(g) = \sum_{a\in\Omega}\frac 12[1-\cos(\frac{2\pi}N\sigma_a)]$, which is just the total string length for $N=2$.

Performing a $S$ transformation, we find the overlap
\begin{align}
	&_{\mathcal A}\inpp{\psi_{ab}}{\hat {\mathcal O}_S}{\psi_{\bar a\bar b}}_{\mathcal A} = \frac 1{|G_\bigtriangleup|}\sum_{\alpha\beta=0}^{N-1}\gamma^{(\bar b-a)\alpha-(b+\bar a)\beta}\sum_{g\in G^{\alpha\beta}_\square}\mathcal A^{-\mathcal L(g)}
\end{align}
If we view strings as domain walls of a $\mathbb Z_N$ clock model on square
lattice described by the following Hamiltonian
 $H =\sum_{\langle ij\rangle} 
\frac 12\left[1-\cos\left(\frac{2\pi}N[\sigma_i-\sigma_j]\right)\right],
\sigma_i,\sigma_j=0,1,\cdots, N-1$,
we find that 
$ N\sum_{g\in G^{00}_\square}\mathcal A^{-\mathcal L(g)}
=\sum_{\{\sigma_i\}} e^{- \beta H}$
can be viewed as the partition function of the $\mathbb Z_N$ clock model, where
$\beta=\log(\mathcal A)$.
In the appendix we show that in the disordered phase of the $\mathbb Z_N$ clock model,
\begin{align}
 Z(\beta)=\sum_{\{\sigma_i\}} e^{-\beta H}=e^{L^2\log(N)-f(\beta)L^2+o(L^{-1})}
\end{align}
to all orders in perturbation theory in $\beta$,
where $f(\beta)$ is a function of $\beta$ only.  Since $N\sum_{g\in
G^{\alpha\beta}_\square}\mathcal A^{-\mathcal L(g)}$ can be viewed as the
partition function of the $\mathbb Z_N$ clock model with twisted boundary condition, we
find that
\begin{align}
\left|
\log
\frac{N\sum_{g\in G^{\alpha\beta}_\square}\mathcal A^{-\mathcal L(g)}
}{
N\sum_{g\in G^{00}_\square}\mathcal A^{-\mathcal L(g)}
} \right| < h Le^{-L/\xi},
\end{align}
where $h$ and $\xi$ are $L$ independent constants.
This is because the total free energies of the $\mathbb Z_N$ clock model with twisted
and untwisted boundary condition can only differ by $h Le^{-L/\xi}$ at most.
Putting everything together, we find that
\begin{equation}
_{\mathcal A}\inpp{\psi_{ab}}{\hat {\mathcal O}_S}{\psi_{\bar a\bar b}}_{\mathcal A} = S_{ab,\bar a\bar b}e^{-[\log N+f(\beta) ] L^2 +o(L^{-1})}
\end{equation}
The universal quantity, $S_{ab,\bar a\bar b}$ is protected, to all orders in $\beta$.

\emph{3D Topological States and $SL(3,\mathbb Z)$}:
According to our conjecture \eqref{eq:overlap} there are similar universal quantities in higher dimensions and it would be interesting to consider a simple example in three dimensions.
For example, the mapping class group of the 3-torus is $\texttt{MCG}(T^3) = SL(3,\mathbb Z)$. This group is generated by two elements of the form \cite{trott1962pair}
\begin{equation}\label{eq:SL3ZGenerators}
	\hat{\tilde S} = \begin{pmatrix}
				0 & 1 & 0 \\
				0 & 0 & 1 \\
				1& 0 & 0
			 \end{pmatrix},
			 \qquad
	\hat{\tilde T}= \begin{pmatrix}
				1 & 0 & 0 \\
				1 & 1 & 0 \\
				0 & 0 & 1
			 \end{pmatrix}.
\end{equation}
These matrices act on the unit vectors by $\hat{\tilde S}:(\hat{\bm x},
\hat{\bm y} , \hat{\bm z}) \mapsto (\hat{\bm z}, \hat{\bm x} , \hat{\bm y})$
and similarly $\hat{\tilde T}:(\hat{\bm x}, \hat{\bm y} , \hat{\bm z}) \mapsto
(\hat{\bm x} + \hat{\bm y}, \hat{\bm y} , \hat{\bm z})$. Thus $\tilde S$
corresponds to a rotation, while $\tilde T$ is shear transformation in the
$xy$-plane. In the case of 3D $\mathbb Z_N$ model, we can directly compute
these generators in a basis with well-defined flux in one direction as \cite{MoradiWen} \[\tilde S_{abc,\bar a\bar b\bar c} =\frac
1N\delta_{b,\bar c}e^{\frac{2\pi i}N(\bar ac - a\bar b)},\quad \tilde
T_{abc,\bar a\bar b\bar c} = \delta_{a,\bar a}\delta_{b,\bar b}\delta_{c,\bar
c}e^{\frac{2\pi i}N ab}.\]
These matrices contain information about self and mutual statistics of particle and string excitations above the ground state  \cite{MoradiWen}.

In the 2D limit where one direction is taken to be
very small, the operator creating a non-contractable loop along this direction is now essentially local. By such a local perturbation, one can break the GSD from $N^3$ down to $N^2$. One can directly show that the generators for an $SL(2,\mathbb Z)\subset SL(3,\mathbb Z)$ subgroup exactly reduce to
the 2D $S$ and $T$ matrices.  \cite{MoradiWen}

\emph{Conclusion}:
In this paper we have conjectured a universal wave function overlap
\eqref{eq:overlap} for gapped systems in $d$ dimensions, which give rise to
projective representations of the mapping class group $\texttt{MCG}(M^d)$, for
any manifold $M^d$. These quantities contain more information than the
topological entanglement entropies
\cite{Kitaev:2005dm,Levin:2006zz,2011PhRvB..84s5120G}, and might characterize
the topological order completely, like in two dimensions \cite{Wrig}.
In a following paper \cite{HeMoradiWen}, we will numerically study the overlaps
\eqref{eq:overlap} for simple two-dimensional models and show that the universal quantities are
very robust against perturbations and unambiguously characterize phase
transitions. In \cite{MoradiWen} we study the universal quantities \eqref{eq:overlap} for three-dimensional systems.

This research is supported by NSF Grant No.
DMR-1005541, NSFC 11074140, and NSFC 11274192. It is also supported by the John
Templeton Foundation. Research at Perimeter Institute is supported by the
Government of Canada through Industry Canada and by the Province of Ontario
through the Ministry of Research.

\bibliography{wencross,all,publst,./Bib}

\appendix

\section{Cumulant Expansion}
Consider the following expansion, which only converges in the disordered phase
\begin{align*}
	Z(\beta)&=\sum_{\{\sigma_i\}}e^{-\beta H} = \exp\left(\log Z(\beta)\right)\\
	&= \exp\Big(\log Z(0)+\frac{Z^\prime(0)}{Z(0)}\beta \\&+ \frac 12\left[-\left(\frac{Z^\prime(0)}{Z(0)}\right)^2+\frac{Z^{\prime\prime}(0)}{Z(0)}\right]\beta^2+o(\beta^3)\Big)\\
	&= \exp\left(\log N L^2 +\sum_{n=1}^\infty\frac{(-1)^n}{n!}\mu_n\beta^n\right),
\end{align*}
where $\mu_n$ is the $n$'th cumulant of the moment of $H$. In particular $\mu_1 = \langle H\rangle_0$, $\mu_2 = \langle H^2\rangle_0-\langle H\rangle^2_0$, $\mu_3 = \langle H^3\rangle_0-3\langle H\rangle_0\langle H^2\rangle_0+2\langle H\rangle^3_0$ and so on. All averages are evaluated at $\beta =0$, $\langle\mathcal O\rangle_0 =\frac 1{Z(0)}\sum_{\{\sigma_i\}}\mathcal O$. Since the averages are taken in the extreme disordered limit (infinite temperature), we have $\langle H\rangle_0 = 2L^2\langle E(\sigma_i,\sigma_j)\rangle_N = L^2$, where each of the $2L^2$ bonds contribute with the average energy $\langle E(\sigma_i,\sigma_j)\rangle_N=\frac 12$.

More generally, consider the total energy to the $n$'th power
\[ H^n = \sum_{i_1\dots i_n}\sum_{\alpha_1,\dots,\alpha_n=x,y}E_{i_1}(\alpha_1)\dots E_{i_n}(\alpha_n),\]
where $E_i(\alpha)=E(\sigma_i-\sigma_{i+\alpha})= \frac 12\left[1-\cos\left(\frac{2\pi}N[\sigma_i-\sigma_{i+\alpha}]\right)\right]$. From here we find
\begin{align*}
&\langle H^n\rangle_0 = \frac 1{Z(0)}\sum_{\{\sigma_i\}}H^n =\\
 &\frac 1{Z(0)} \sum_{\{\sigma_i\}}\Biggl(\sum_i\termOne + \sum_{i\neq j} \termTwo +\sum_{i\neq j\neq k} \termThree +\dots\Biggl),
\end{align*}
where $\termOneSmall$ contain all possible terms where all $E_i(\alpha)$ are at site $i$, while $\termTwoSmall$ contain all possible terms where all $E_i(\alpha)$ are at $i$ and $j$, and so on. The first term is given by
\[\sum_i\termOne = 
\sum_i\sum_{q=0}^n
\begin{pmatrix}
n\\q
\end{pmatrix}E_i^q(x)E_i^{n-q}(y)\equiv 
\sum_i\mathcal M_i(n).
\]
Similarly, the next two terms are given by
\[ \sum_{i_1\neq i_2}\termTwoTwo = \frac 12\sum_{i_1\neq i_2}\sum_{k=1}^{n-1}\begin{pmatrix}
n\\k
\end{pmatrix}\mathcal M_{i_1}(k)\mathcal M_{i_2}(n-k),	\]
and 
\begin{align*}
 \sum_{i_1\neq i_2\neq i_3}\termThreeTwo &= \frac 1{3!}\sum_{i_1\neq i_2\neq i_3}\sum_{k_1+k_2+k_3=n}^{1\leq k_i\leq n-2}\begin{pmatrix}
n\\k_1,k_2,k_3
\end{pmatrix}\\
&\qquad\times\mathcal M_{i_1}(k_1)\mathcal M_{i_2}(k_2)\mathcal M_{i_3}(k_3),
\end{align*}
where the symbol $\begin{pmatrix}
 	n\\k_1,k_2,\dots, k_m
 \end{pmatrix} = \frac{n!}{k_1!k_2!\dots k_m!}$ is the multinomial coefficient. One can verify that in general the expectation value takes the form
\begin{align*}
	\langle H^n\rangle_0&=\frac 1{Z(0)}\sum_{\{\sigma_i\}}\sum_{p=1}^n\frac 1{p!}\sum_{i_1\neq\dots\neq i_p}\sum_{k_1+\dots + k_p=n}^{1\leq k_i\leq n-p+1}\begin{pmatrix}
n\\k_1,\dots,k_p
\end{pmatrix}\\
&\qquad\times\mathcal M_{i_1}(k_1)\dots\mathcal M_{i_p}(k_p)\\
&= \frac 1{Z(0)}\sum_{p=1}^n\frac 1{p!}\sum_{i_1\neq\dots\neq i_p}\sum_{\{\sigma\}_{i_1,\dots,i_p}}\sum_{k_1+\dots + k_p=n}^{1\leq k_i\leq n-p+1}\begin{pmatrix}
n\\k_1,\dots,k_p
\end{pmatrix}\\
&\times\Bigg[\hspace{20pt}\sum_{\mathclap{\sigma_{i_1}\sigma_{i_1+x}\sigma_{i_1+y}}}\mathcal M_{i_1}(k_1)\Bigg]\dots \Bigg[\hspace{20pt}\sum_{\mathclap{\sigma_{i_p}\sigma_{i_p+x}\sigma_{i_p+y}}}\mathcal M_{i_p}(k_p)\Bigg].
\end{align*}
The notation $\{\sigma\}_{i_1,\dots,i_p}$ stands for the set of $\sigma_i$ for all $i$, except $\sigma_{i_1},\sigma_{i_1+x},\sigma_{i_1+y},\dots, \sigma_{i_p},\sigma_{i_p+x},\sigma_{i_p+y}$. This sum is therefore $\sum_{\{\sigma\}_{i_1,\dots,i_p}}1 = N^{L^2-3p}$. The other sum can be rewritten as follows
\begin{align*}
	\sum_{\mathclap{\sigma_{i}\sigma_{i+x}\sigma_{i+y}}}\mathcal M_{i}(k) &= \sum_{q=0}^k
\begin{pmatrix}
k\\q
\end{pmatrix}\hspace{10pt}\sum_{\mathclap{\sigma_{i}\sigma_{i+x}\sigma_{i+y}}}E^q(\sigma_i-\sigma_{i+x})E^{k-q}(\sigma_i-\sigma_{i+y})\\
&= N^3\sum_{q=0}^k
\begin{pmatrix}
k\\q
\end{pmatrix}\langle E^q\rangle_N \langle E^{k-q}\rangle_N \equiv N^3\tilde{\mathcal M}(k).
\end{align*}
Here we have defined the link-averages $\langle E^q\rangle_N \equiv \frac 1N\sum_AE^q(A)$ and the second line follows from changing variables in the $\sigma$ sums. The $N^3$ factors together with the fact that $Z(0)=N^{L^2}$ cancel the $N^{L^2-3p}$ factor. Now, since $\tilde{\mathcal M}(k)$ does not depend on the site $i$ any more we get another factor of $\sum_{i_1\neq\dots\neq i_p}1 = (L^2)_p$, where
\[ (L^2)_p  = \frac{\Gamma(L^2+1)}{\Gamma(L^2-p+1)} = L^2(L^2-1)\dots (L^2-p+1)		\]
is the descending Pochhammer symbol.
Collecting all this together, we find the following simplified expression for the total energy moments
\begin{equation}\label{eq:HnAverage}
	\langle H^n\rangle_0	= \sum_{p=1}^nC_p(n) (L^2)_p,
\end{equation}
with the coefficients
\begin{align*} C_p(n) &= \frac 1{p!}\sum_{k_1+\dots + k_p=n}^{1\leq k_i\leq n-p+1}\begin{pmatrix}
 	n\\k_1\dots, k_p
 \end{pmatrix} \tilde{\mathcal M}(k_1)\dots\tilde{\mathcal M}(k_p).
\end{align*}
The usefulness of \eqref{eq:HnAverage} comes from the fact that the $L^2$ dependence is explicitly factorized.

We note that the total energy moments in equation \eqref{eq:HnAverage} goes as $o(L^2)$ in the volume, and in particular do not contain any constant terms. The cumulants $\mu_n$ are just sums and differences of these moments, and are therefore of order $o(L^2)$. Thus the cumulants do not have any constant terms at all.

Furthermore, it is well known from statistical physics that the free energy $F(\beta)=\log Z(\beta)$ is an extensive quantity and scales as volume in the thermodynamic limit. This implies that all the moments also scale as $L^2$ and the higher order terms must cancel out (this can directly be verified
for the first few moments using \eqref{eq:HnAverage}). Thus we conclude that
\begin{equation}
	Z(\beta) = e^{\log N L^2 - f_N(\beta)L^2 + o(L^{-1})},
\end{equation}
in particular there is no constant term to all orders of $\beta$. As a few examples, we can use \eqref{eq:HnAverage} to calculate
\begin{align*}
	f_2(\beta) &= \beta -\frac{\beta^2}{4}+\frac{\beta^4}{96}-\frac{\beta^6}{1440}+o(\beta^7),\\
	f_3(\beta) &= \beta -\frac{\beta^2}{8}-\frac{\beta^3}{96}+\frac{\beta^4}{512}+\frac{\beta^5}{2048}-\frac{7 \beta^6}{245760}+o(\beta^7),\\
		f_4(\beta) &= \beta -\frac{\beta^2}{8}+\frac{\beta^4}{768}-\frac{\beta^6}{46080}+o(\beta^7).
\end{align*}

\end{document}

%% file: tikzDef.tex
\newcommand{\termOne}{
\begin{tikzpicture}[scale=0.9,baseline={([yshift=-.5ex]current bounding box.center)}]
\draw[step=0.3,black,thin] (0,0) grid (1.2,1.2);
\draw[red, ultra thick] (0.3,0.9) -- (0.6,0.9);
\draw[red, ultra thick] (0.3,0.9) -- (0.3,1.2);
\node at (0.45,1.05) {\scalebox{0.7}{$i$}};
\end{tikzpicture}
}

\newcommand{\termTwo}{
\begin{tikzpicture}[scale=0.9,baseline={([yshift=-.5ex]current bounding box.center)}]
\draw[step=0.3,black,thin] (0,0) grid (1.2,1.2);

\draw[red, ultra thick] (0.3,0.9) -- (0.6,0.9);
\draw[red, ultra thick] (0.3,0.9) -- (0.3,1.2);

\draw[blue, ultra thick] (0.9,0.3) -- (1.2,0.3);
\draw[blue, ultra thick] (0.9,0.3) -- (0.9,0.6);

\node at (0.45,1.05) {\scalebox{0.7}{$i$}};
\node at (1.05,0.45) {\scalebox{0.7}{$j$}};
\end{tikzpicture}
}

\newcommand{\termThree}{
\begin{tikzpicture}[scale=0.9,baseline={([yshift=-.5ex]current bounding box.center)}]
\draw[step=0.3,black,thin] (0,0) grid (1.2,1.2);

\draw[red, ultra thick] (0.3,0.9) -- (0.6,0.9);
\draw[red, ultra thick] (0.3,0.9) -- (0.3,1.2);

\draw[blue, ultra thick] (0.9,0.3) -- (1.2,0.3);
\draw[blue, ultra thick] (0.9,0.3) -- (0.9,0.6);

\draw[green, ultra thick] (0.0,0.0) -- (0.3,0.0);
\draw[green, ultra thick] (0.0,0.0) -- (0.0,0.3);

\node at (0.45,1.05) {\scalebox{0.7}{$i$}};
\node at (1.05,0.45) {\scalebox{0.7}{$j$}};
\node at (0.15,0.15) {\scalebox{0.7}{$k$}};
\end{tikzpicture}
}

\newcommand{\termOneSmall}{
\begin{tikzpicture}[scale=0.5,baseline={([yshift=-.5ex]current bounding box.center)}]
\draw[step=0.3,black,thin] (0,0) grid (1.2,1.2);
\draw[red, ultra thick] (0.3,0.9) -- (0.6,0.9);
\draw[red, ultra thick] (0.3,0.9) -- (0.3,1.2);
\node at (0.45,1.05) {\scalebox{0.4}{$i$}};
\end{tikzpicture}
}

\newcommand{\termTwoSmall}{
\begin{tikzpicture}[scale=0.5,baseline={([yshift=-.5ex]current bounding box.center)}]
\draw[step=0.3,black,thin] (0,0) grid (1.2,1.2);

\draw[red, ultra thick] (0.3,0.9) -- (0.6,0.9);
\draw[red, ultra thick] (0.3,0.9) -- (0.3,1.2);

\draw[blue, ultra thick] (0.9,0.3) -- (1.2,0.3);
\draw[blue, ultra thick] (0.9,0.3) -- (0.9,0.6);

\node at (0.45,1.05) {\scalebox{0.4}{$i$}};
\node at (1.05,0.45) {\scalebox{0.4}{$j$}};
\end{tikzpicture}
}

\newcommand{\termTwoTwo}{
\begin{tikzpicture}[scale=0.9,baseline={([yshift=-.5ex]current bounding box.center)}]
\draw[step=0.3,black,thin] (0,0) grid (1.2,1.2);

\draw[red, ultra thick] (0.3,0.9) -- (0.6,0.9);
\draw[red, ultra thick] (0.3,0.9) -- (0.3,1.2);

\draw[blue, ultra thick] (0.9,0.3) -- (1.2,0.3);
\draw[blue, ultra thick] (0.9,0.3) -- (0.9,0.6);

\node at (0.465,1.05) {\scalebox{0.7}{$i_1$}};
\node at (1.065,0.45) {\scalebox{0.7}{$i_2$}};
\end{tikzpicture}
}

\newcommand{\termThreeTwo}{
\begin{tikzpicture}[scale=0.9,baseline={([yshift=-.5ex]current bounding box.center)}]
\draw[step=0.3,black,thin] (0,0) grid (1.2,1.2);

\draw[red, ultra thick] (0.3,0.9) -- (0.6,0.9);
\draw[red, ultra thick] (0.3,0.9) -- (0.3,1.2);

\draw[blue, ultra thick] (0.9,0.3) -- (1.2,0.3);
\draw[blue, ultra thick] (0.9,0.3) -- (0.9,0.6);

\draw[green, ultra thick] (0.0,0.0) -- (0.3,0.0);
\draw[green, ultra thick] (0.0,0.0) -- (0.0,0.3);

\node at (0.465,1.05) {\scalebox{0.7}{$i_1$}};
\node at (1.065,0.45) {\scalebox{0.7}{$i_2$}};
\node at (0.165,0.15) {\scalebox{0.7}{$i_3$}};
\end{tikzpicture}
}

%% file: STmat.bbl
\begin{thebibliography}{44}%
\makeatletter
\providecommand \@ifxundefined [1]{%
 \@ifx{#1\undefined}
}%
\providecommand \@ifnum [1]{%
 \ifnum #1\expandafter \@firstoftwo
 \else \expandafter \@secondoftwo
 \fi
}%
\providecommand \@ifx [1]{%
 \ifx #1\expandafter \@firstoftwo
 \else \expandafter \@secondoftwo
 \fi
}%
\providecommand \natexlab [1]{#1}%
\providecommand \enquote  [1]{``#1''}%
\providecommand \bibnamefont  [1]{#1}%
\providecommand \bibfnamefont [1]{#1}%
\providecommand \citenamefont [1]{#1}%
\providecommand \href@noop [0]{\@secondoftwo}%
\providecommand \href [0]{\begingroup \@sanitize@url \@href}%
\providecommand \@href[1]{\@@startlink{#1}\@@href}%
\providecommand \@@href[1]{\endgroup#1\@@endlink}%
\providecommand \@sanitize@url [0]{\catcode `\\12\catcode `\$12\catcode
  `\&12\catcode `\#12\catcode `\^12\catcode `\_12\catcode `\%12\relax}%
\providecommand \@@startlink[1]{}%
\providecommand \@@endlink[0]{}%
\providecommand \url  [0]{\begingroup\@sanitize@url \@url }%
\providecommand \@url [1]{\endgroup\@href {#1}{\urlprefix }}%
\providecommand \urlprefix  [0]{URL }%
\providecommand \Eprint [0]{\href }%
\providecommand \doibase [0]{http://dx.doi.org/}%
\providecommand \selectlanguage [0]{\@gobble}%
\providecommand \bibinfo  [0]{\@secondoftwo}%
\providecommand \bibfield  [0]{\@secondoftwo}%
\providecommand \translation [1]{[#1]}%
\providecommand \BibitemOpen [0]{}%
\providecommand \bibitemStop [0]{}%
\providecommand \bibitemNoStop [0]{.\EOS\space}%
\providecommand \EOS [0]{\spacefactor3000\relax}%
\providecommand \BibitemShut  [1]{\csname bibitem#1\endcsname}%
\let\auto@bib@innerbib\@empty
%</preamble>
\bibitem [{\citenamefont {Tsui}\ \emph {et~al.}(1982)\citenamefont {Tsui},
  \citenamefont {Stormer},\ and\ \citenamefont {Gossard}}]{TSG8259}%
  \BibitemOpen
  \bibfield  {author} {\bibinfo {author} {\bibfnamefont {D.~C.}\ \bibnamefont
  {Tsui}}, \bibinfo {author} {\bibfnamefont {H.~L.}\ \bibnamefont {Stormer}}, \
  and\ \bibinfo {author} {\bibfnamefont {A.~C.}\ \bibnamefont {Gossard}},\
  }\href@noop {} {\bibfield  {journal} {\bibinfo  {journal} {Phys. Rev. Lett.}\
  }\textbf {\bibinfo {volume} {48}},\ \bibinfo {pages} {1559} (\bibinfo {year}
  {1982})}\BibitemShut {NoStop}%
\bibitem [{\citenamefont {Laughlin}(1983)}]{L8395}%
  \BibitemOpen
  \bibfield  {author} {\bibinfo {author} {\bibfnamefont {R.~B.}\ \bibnamefont
  {Laughlin}},\ }\href@noop {} {\bibfield  {journal} {\bibinfo  {journal}
  {Phys. Rev. Lett.}\ }\textbf {\bibinfo {volume} {50}},\ \bibinfo {pages}
  {1395} (\bibinfo {year} {1983})}\BibitemShut {NoStop}%
\bibitem [{\citenamefont {Kalmeyer}\ and\ \citenamefont
  {Laughlin}(1987)}]{KL8795}%
  \BibitemOpen
  \bibfield  {author} {\bibinfo {author} {\bibfnamefont {V.}~\bibnamefont
  {Kalmeyer}}\ and\ \bibinfo {author} {\bibfnamefont {R.~B.}\ \bibnamefont
  {Laughlin}},\ }\href@noop {} {\bibfield  {journal} {\bibinfo  {journal}
  {Phys. Rev. Lett.}\ }\textbf {\bibinfo {volume} {59}},\ \bibinfo {pages}
  {2095} (\bibinfo {year} {1987})}\BibitemShut {NoStop}%
\bibitem [{\citenamefont {Wen}\ \emph {et~al.}(1989)\citenamefont {Wen},
  \citenamefont {Wilczek},\ and\ \citenamefont {Zee}}]{WWZcsp}%
  \BibitemOpen
  \bibfield  {author} {\bibinfo {author} {\bibfnamefont {X.-G.}\ \bibnamefont
  {Wen}}, \bibinfo {author} {\bibfnamefont {F.}~\bibnamefont {Wilczek}}, \ and\
  \bibinfo {author} {\bibfnamefont {A.}~\bibnamefont {Zee}},\ }\href@noop {}
  {\bibfield  {journal} {\bibinfo  {journal} {Phys. Rev. B}\ }\textbf {\bibinfo
  {volume} {39}},\ \bibinfo {pages} {11413} (\bibinfo {year}
  {1989})}\BibitemShut {NoStop}%
\bibitem [{\citenamefont {Wen}(1989)}]{Wtop}%
  \BibitemOpen
  \bibfield  {author} {\bibinfo {author} {\bibfnamefont {X.-G.}\ \bibnamefont
  {Wen}},\ }\href@noop {} {\bibfield  {journal} {\bibinfo  {journal} {Phys.
  Rev. B}\ }\textbf {\bibinfo {volume} {40}},\ \bibinfo {pages} {7387}
  (\bibinfo {year} {1989})}\BibitemShut {NoStop}%
\bibitem [{\citenamefont {Wen}\ and\ \citenamefont {Niu}(1990)}]{WNtop}%
  \BibitemOpen
  \bibfield  {author} {\bibinfo {author} {\bibfnamefont {X.-G.}\ \bibnamefont
  {Wen}}\ and\ \bibinfo {author} {\bibfnamefont {Q.}~\bibnamefont {Niu}},\
  }\href@noop {} {\bibfield  {journal} {\bibinfo  {journal} {Phys. Rev. B}\
  }\textbf {\bibinfo {volume} {41}},\ \bibinfo {pages} {9377} (\bibinfo {year}
  {1990})}\BibitemShut {NoStop}%
\bibitem [{\citenamefont {Wen}(1990)}]{Wrig}%
  \BibitemOpen
  \bibfield  {author} {\bibinfo {author} {\bibfnamefont {X.-G.}\ \bibnamefont
  {Wen}},\ }\href@noop {} {\bibfield  {journal} {\bibinfo  {journal} {Int. J.
  Mod. Phys. B}\ }\textbf {\bibinfo {volume} {4}},\ \bibinfo {pages} {239}
  (\bibinfo {year} {1990})}\BibitemShut {NoStop}%
\bibitem [{\citenamefont {Keski-Vakkuri}\ and\ \citenamefont
  {Wen}(1993)}]{KW9327}%
  \BibitemOpen
  \bibfield  {author} {\bibinfo {author} {\bibfnamefont {E.}~\bibnamefont
  {Keski-Vakkuri}}\ and\ \bibinfo {author} {\bibfnamefont {X.-G.}\ \bibnamefont
  {Wen}},\ }\href@noop {} {\bibfield  {journal} {\bibinfo  {journal} {Int. J.
  Mod. Phys. B}\ }\textbf {\bibinfo {volume} {7}},\ \bibinfo {pages} {4227}
  (\bibinfo {year} {1993})}\BibitemShut {NoStop}%
\bibitem [{\citenamefont {Wen}(2012)}]{W1221}%
  \BibitemOpen
  \bibfield  {author} {\bibinfo {author} {\bibfnamefont {X.-G.}\ \bibnamefont
  {Wen}},\ }\href@noop {} {\  (\bibinfo {year} {2012})},\ \Eprint
  {http://arxiv.org/abs/arXiv:1212.5121} {arXiv:1212.5121} \BibitemShut
  {NoStop}%
\bibitem [{\citenamefont {Chen}\ \emph {et~al.}(2010)\citenamefont {Chen},
  \citenamefont {Gu},\ and\ \citenamefont {Wen}}]{CGW1038}%
  \BibitemOpen
  \bibfield  {author} {\bibinfo {author} {\bibfnamefont {X.}~\bibnamefont
  {Chen}}, \bibinfo {author} {\bibfnamefont {Z.-C.}\ \bibnamefont {Gu}}, \ and\
  \bibinfo {author} {\bibfnamefont {X.-G.}\ \bibnamefont {Wen}},\ }\href@noop
  {} {\bibfield  {journal} {\bibinfo  {journal} {Phys. Rev. B}\ }\textbf
  {\bibinfo {volume} {82}},\ \bibinfo {pages} {155138} (\bibinfo {year}
  {2010})},\ \Eprint {http://arxiv.org/abs/arXiv:1004.3835} {arXiv:1004.3835}
  \BibitemShut {NoStop}%
\bibitem [{\citenamefont {Wang}(2010)}]{Wang10}%
  \BibitemOpen
  \bibfield  {author} {\bibinfo {author} {\bibfnamefont {Z.}~\bibnamefont
  {Wang}},\ }\href@noop {} {\emph {\bibinfo {title} {Topological Quantum
  Computation}}}\ (\bibinfo  {publisher} {CBMS Regional Conference Series in
  Mathematics},\ \bibinfo {year} {2010})\BibitemShut {NoStop}%
\bibitem [{\citenamefont {Wen}(1992)}]{Wedgerev}%
  \BibitemOpen
  \bibfield  {author} {\bibinfo {author} {\bibfnamefont {X.-G.}\ \bibnamefont
  {Wen}},\ }\href@noop {} {\bibfield  {journal} {\bibinfo  {journal} {Int. J.
  Mod. Phys. B}\ }\textbf {\bibinfo {volume} {6}},\ \bibinfo {pages} {1711}
  (\bibinfo {year} {1992})}\BibitemShut {NoStop}%
\bibitem [{\citenamefont {Wen}(1995)}]{Wtoprev}%
  \BibitemOpen
  \bibfield  {author} {\bibinfo {author} {\bibfnamefont {X.-G.}\ \bibnamefont
  {Wen}},\ }\href@noop {} {\bibfield  {journal} {\bibinfo  {journal} {Advances
  in Physics}\ }\textbf {\bibinfo {volume} {44}},\ \bibinfo {pages} {405}
  (\bibinfo {year} {1995})}\BibitemShut {NoStop}%
\bibitem [{\citenamefont {Zhang}\ \emph {et~al.}(2012)\citenamefont {Zhang},
  \citenamefont {Grover}, \citenamefont {Turner}, \citenamefont {Oshikawa},\
  and\ \citenamefont {Vishwanath}}]{ZGT1251}%
  \BibitemOpen
  \bibfield  {author} {\bibinfo {author} {\bibfnamefont {Y.}~\bibnamefont
  {Zhang}}, \bibinfo {author} {\bibfnamefont {T.}~\bibnamefont {Grover}},
  \bibinfo {author} {\bibfnamefont {A.}~\bibnamefont {Turner}}, \bibinfo
  {author} {\bibfnamefont {M.}~\bibnamefont {Oshikawa}}, \ and\ \bibinfo
  {author} {\bibfnamefont {A.}~\bibnamefont {Vishwanath}},\ }\href {\doibase
  10.1103/PhysRevB.85.235151} {\bibfield  {journal} {\bibinfo  {journal} {Phys.
  Rev. B}\ }\textbf {\bibinfo {volume} {85}},\ \bibinfo {pages} {235151}
  (\bibinfo {year} {2012})},\ \Eprint {http://arxiv.org/abs/arXiv:1111.2342}
  {arXiv:1111.2342} \BibitemShut {NoStop}%
\bibitem [{\citenamefont {Cincio}\ and\ \citenamefont {Vidal}(2013)}]{CV1223}%
  \BibitemOpen
  \bibfield  {author} {\bibinfo {author} {\bibfnamefont {L.}~\bibnamefont
  {Cincio}}\ and\ \bibinfo {author} {\bibfnamefont {G.}~\bibnamefont {Vidal}},\
  }\href@noop {} {\bibfield  {journal} {\bibinfo  {journal} {Phys. Rev. Lett.}\
  }\textbf {\bibinfo {volume} {110}},\ \bibinfo {pages} {067208} (\bibinfo
  {year} {2013})},\ \Eprint {http://arxiv.org/abs/arXiv:1208.2623}
  {arXiv:1208.2623} \BibitemShut {NoStop}%
\bibitem [{\citenamefont {Zaletel}\ \emph {et~al.}(2012)\citenamefont
  {Zaletel}, \citenamefont {Mong},\ and\ \citenamefont {Pollmann}}]{ZMP1233}%
  \BibitemOpen
  \bibfield  {author} {\bibinfo {author} {\bibfnamefont {M.~P.}\ \bibnamefont
  {Zaletel}}, \bibinfo {author} {\bibfnamefont {R.~S.~K.}\ \bibnamefont
  {Mong}}, \ and\ \bibinfo {author} {\bibfnamefont {F.}~\bibnamefont
  {Pollmann}},\ }\href@noop {} {\  (\bibinfo {year} {2012})},\ \Eprint
  {http://arxiv.org/abs/arXiv:1211.3733} {arXiv:1211.3733} \BibitemShut
  {NoStop}%
\bibitem [{\citenamefont {Tu}\ \emph {et~al.}(2012)\citenamefont {Tu},
  \citenamefont {Zhang},\ and\ \citenamefont {Qi}}]{TZQ1251}%
  \BibitemOpen
  \bibfield  {author} {\bibinfo {author} {\bibfnamefont {H.-H.}\ \bibnamefont
  {Tu}}, \bibinfo {author} {\bibfnamefont {Y.}~\bibnamefont {Zhang}}, \ and\
  \bibinfo {author} {\bibfnamefont {X.-L.}\ \bibnamefont {Qi}},\ }\href
  {http://arxiv.org/abs/1212.6951} {\bibfield  {journal} {\bibinfo  {journal}
  {arXiv preprint arXiv:1212.6951}\ } (\bibinfo {year} {2012})}\BibitemShut
  {NoStop}%
\bibitem [{\citenamefont {Kitaev}\ and\ \citenamefont
  {Preskill}(2006)}]{Kitaev:2005dm}%
  \BibitemOpen
  \bibfield  {author} {\bibinfo {author} {\bibfnamefont {A.}~\bibnamefont
  {Kitaev}}\ and\ \bibinfo {author} {\bibfnamefont {J.}~\bibnamefont
  {Preskill}},\ }\href {\doibase 10.1103/PhysRevLett.96.110404} {\bibfield
  {journal} {\bibinfo  {journal} {Phys.Rev.Lett.}\ }\textbf {\bibinfo {volume}
  {96}},\ \bibinfo {pages} {110404} (\bibinfo {year} {2006})},\ \Eprint
  {http://arxiv.org/abs/hep-th/0510092} {arXiv:hep-th/0510092 [hep-th]}
  \BibitemShut {NoStop}%
%%CITATION = HEP-TH/0510092;%%
\bibitem [{\citenamefont {Levin}\ and\ \citenamefont
  {Wen}(2006)}]{Levin:2006zz}%
  \BibitemOpen
  \bibfield  {author} {\bibinfo {author} {\bibfnamefont {M.}~\bibnamefont
  {Levin}}\ and\ \bibinfo {author} {\bibfnamefont {X.-G.}\ \bibnamefont
  {Wen}},\ }\href {\doibase 10.1103/PhysRevLett.96.110405} {\bibfield
  {journal} {\bibinfo  {journal} {Phys.Rev.Lett.}\ }\textbf {\bibinfo {volume}
  {96}},\ \bibinfo {pages} {110405} (\bibinfo {year} {2006})},\ \Eprint
  {http://arxiv.org/abs/cond-mat/0510613} {arXiv:cond-mat/0510613 [cond-mat]}
  \BibitemShut {NoStop}%
%%CITATION = PRLTA,96,110405;%%
\bibitem [{\citenamefont {{Grover}}\ \emph {et~al.}(2011)\citenamefont
  {{Grover}}, \citenamefont {{Turner}},\ and\ \citenamefont
  {{Vishwanath}}}]{2011PhRvB..84s5120G}%
  \BibitemOpen
  \bibfield  {author} {\bibinfo {author} {\bibfnamefont {T.}~\bibnamefont
  {{Grover}}}, \bibinfo {author} {\bibfnamefont {A.~M.}\ \bibnamefont
  {{Turner}}}, \ and\ \bibinfo {author} {\bibfnamefont {A.}~\bibnamefont
  {{Vishwanath}}},\ }\href {\doibase 10.1103/PhysRevB.84.195120} {\bibfield
  {journal} {\bibinfo  {journal} {\prb}\ }\textbf {\bibinfo {volume} {84}},\
  \bibinfo {eid} {195120} (\bibinfo {year} {2011})},\ \Eprint
  {http://arxiv.org/abs/1108.4038} {arXiv:1108.4038 [cond-mat.str-el]}
  \BibitemShut {NoStop}%
\bibitem [{\citenamefont {Hung}\ and\ \citenamefont {Wen}(2013)}]{HW1339}%
  \BibitemOpen
  \bibfield  {author} {\bibinfo {author} {\bibfnamefont {L.-Y.}\ \bibnamefont
  {Hung}}\ and\ \bibinfo {author} {\bibfnamefont {X.-G.}\ \bibnamefont {Wen}},\
  }\href@noop {} {\  (\bibinfo {year} {2013})},\ \Eprint
  {http://arxiv.org/abs/arXiv:1311.5539} {arXiv:1311.5539} \BibitemShut
  {NoStop}%
\bibitem [{\citenamefont {Kong}\ and\ \citenamefont {Wen}(2014)}]{KW13}%
  \BibitemOpen
  \bibfield  {author} {\bibinfo {author} {\bibfnamefont {L.}~\bibnamefont
  {Kong}}\ and\ \bibinfo {author} {\bibfnamefont {X.-G.}\ \bibnamefont {Wen}},\
  }\href@noop {} {\bibfield  {journal} {\bibinfo  {journal} {to appear}\ }
  (\bibinfo {year} {2014})}\BibitemShut {NoStop}%
\bibitem [{\citenamefont {Read}\ and\ \citenamefont {Sachdev}(1991)}]{RS9173}%
  \BibitemOpen
  \bibfield  {author} {\bibinfo {author} {\bibfnamefont {N.}~\bibnamefont
  {Read}}\ and\ \bibinfo {author} {\bibfnamefont {S.}~\bibnamefont {Sachdev}},\
  }\href@noop {} {\bibfield  {journal} {\bibinfo  {journal} {Phys. Rev. Lett.}\
  }\textbf {\bibinfo {volume} {66}},\ \bibinfo {pages} {1773} (\bibinfo {year}
  {1991})}\BibitemShut {NoStop}%
\bibitem [{\citenamefont {Wen}(1991)}]{Wsrvb}%
  \BibitemOpen
  \bibfield  {author} {\bibinfo {author} {\bibfnamefont {X.-G.}\ \bibnamefont
  {Wen}},\ }\href@noop {} {\bibfield  {journal} {\bibinfo  {journal} {Phys.
  Rev. B}\ }\textbf {\bibinfo {volume} {44}},\ \bibinfo {pages} {2664}
  (\bibinfo {year} {1991})}\BibitemShut {NoStop}%
\bibitem [{\citenamefont {Bullock}\ and\ \citenamefont
  {Brennen}(2007)}]{BB0781}%
  \BibitemOpen
  \bibfield  {author} {\bibinfo {author} {\bibfnamefont {S.~S.}\ \bibnamefont
  {Bullock}}\ and\ \bibinfo {author} {\bibfnamefont {G.~K.}\ \bibnamefont
  {Brennen}},\ }\href {\doibase 10.1088/1751-8113/40/13/013} {\bibfield
  {journal} {\bibinfo  {journal} {J. Phys. A}\ }\textbf {\bibinfo {volume}
  {A40}},\ \bibinfo {pages} {3481} (\bibinfo {year} {2007})},\ \Eprint
  {http://arxiv.org/abs/quant-ph/0609070} {arXiv:quant-ph/0609070 [quant-ph]}
  \BibitemShut {NoStop}%
%%CITATION = QUANT-PH/0609070;%%
\bibitem [{\citenamefont {{Schulz}}\ \emph {et~al.}(2012)\citenamefont
  {{Schulz}}, \citenamefont {{Dusuel}}, \citenamefont {{Or{\'u}s}},
  \citenamefont {{Vidal}},\ and\ \citenamefont {{Schmidt}}}]{SchulzetAlZn}%
  \BibitemOpen
  \bibfield  {author} {\bibinfo {author} {\bibfnamefont {M.~D.}\ \bibnamefont
  {{Schulz}}}, \bibinfo {author} {\bibfnamefont {S.}~\bibnamefont {{Dusuel}}},
  \bibinfo {author} {\bibfnamefont {R.}~\bibnamefont {{Or{\'u}s}}}, \bibinfo
  {author} {\bibfnamefont {J.}~\bibnamefont {{Vidal}}}, \ and\ \bibinfo
  {author} {\bibfnamefont {K.~P.}\ \bibnamefont {{Schmidt}}},\ }\href {\doibase
  10.1088/1367-2630/14/2/025005} {\bibfield  {journal} {\bibinfo  {journal}
  {New Journal of Physics}\ }\textbf {\bibinfo {volume} {14}},\ \bibinfo {eid}
  {025005} (\bibinfo {year} {2012})},\ \Eprint {http://arxiv.org/abs/1110.3632}
  {arXiv:1110.3632 [cond-mat.stat-mech]} \BibitemShut {NoStop}%
\bibitem [{\citenamefont {You}\ and\ \citenamefont {Wen}(2012)}]{YW1207}%
  \BibitemOpen
  \bibfield  {author} {\bibinfo {author} {\bibfnamefont {Y.-Z.}\ \bibnamefont
  {You}}\ and\ \bibinfo {author} {\bibfnamefont {X.-G.}\ \bibnamefont {Wen}},\
  }\href@noop {} {\bibfield  {journal} {\bibinfo  {journal} {Phys. Rev. B}\
  }\textbf {\bibinfo {volume} {86}},\ \bibinfo {pages} {161107} (\bibinfo
  {year} {2012})},\ \Eprint {http://arxiv.org/abs/arXiv:1204.0113}
  {arXiv:1204.0113} \BibitemShut {NoStop}%
\bibitem [{Note1()}]{Note1}%
  \BibitemOpen
  \bibinfo {note} {The ambiguity of $c_-$ can be understood by the existence of
  the so-called $E_8$ state, which can be realized by a Chern-Simons theory
  where the $K$ matrix is the Cartan matrix of $E_8$. This theory has only
  trivial bulk excitations since $\protect \qopname \relax m{det}K=1$ but
  boundary theory given by the affine Lie algebra $(\protect \mathaccentV
  {hat}05EE_8)_1$, which has $c_- =8$. Thus there is always the ambiguity of
  adding a $E_8$ state without changing the bulk excitations, but shifting the
  chiral central charge by $8$. The chiral central charge is related to
  perturbative gravitational anomalies on the edge, which signals lack of
  energy conservation, or a gravitational parity anomaly from the bulk
  perspective. Physically, this corresponds to a thermal Hall effect by the
  Callan-Harvey inflow mechanism \cite {Callan:1984sa} and is a consequence of
  the decent relations of anomalies in different dimensions. Note that in the
  case $c_- = 0$, the edge states are not chiral but they can however still be
  completely robust. \cite {L1309} This is related to global gravitational
  anomalies, ie modular anomaly on the edge.}\BibitemShut {Stop}%
\bibitem [{\citenamefont {{Hastings}}(2007)}]{Hastings2007}%
  \BibitemOpen
  \bibfield  {author} {\bibinfo {author} {\bibfnamefont {M.~B.}\ \bibnamefont
  {{Hastings}}},\ }\href {\doibase 10.1088/1742-5468/2007/08/P08024} {\bibfield
   {journal} {\bibinfo  {journal} {Journal of Statistical Mechanics: Theory and
  Experiment}\ }\textbf {\bibinfo {volume} {8}},\ \bibinfo {pages} {24}
  (\bibinfo {year} {2007})},\ \Eprint {http://arxiv.org/abs/0705.2024}
  {arXiv:0705.2024 [quant-ph]} \BibitemShut {NoStop}%
\bibitem [{\citenamefont {Masanes}(2009)}]{Masanes:2009tg}%
  \BibitemOpen
  \bibfield  {author} {\bibinfo {author} {\bibfnamefont {L.}~\bibnamefont
  {Masanes}},\ }\href {\doibase 10.1103/PhysRevA.80.052104} {\bibfield
  {journal} {\bibinfo  {journal} {Phys.Rev.}\ }\textbf {\bibinfo {volume}
  {A80}},\ \bibinfo {pages} {052104} (\bibinfo {year} {2009})},\ \Eprint
  {http://arxiv.org/abs/0907.4672} {arXiv:0907.4672 [quant-ph]} \BibitemShut
  {NoStop}%
%%CITATION = ARXIV:0907.4672;%%
\bibitem [{\citenamefont {{Chen}}\ \emph {et~al.}(2010)\citenamefont {{Chen}},
  \citenamefont {{Zeng}}, \citenamefont {{Gu}}, \citenamefont {{Chuang}},\ and\
  \citenamefont {{Wen}}}]{2010PhRvB..82p5119C}%
  \BibitemOpen
  \bibfield  {author} {\bibinfo {author} {\bibfnamefont {X.}~\bibnamefont
  {{Chen}}}, \bibinfo {author} {\bibfnamefont {B.}~\bibnamefont {{Zeng}}},
  \bibinfo {author} {\bibfnamefont {Z.-C.}\ \bibnamefont {{Gu}}}, \bibinfo
  {author} {\bibfnamefont {I.~L.}\ \bibnamefont {{Chuang}}}, \ and\ \bibinfo
  {author} {\bibfnamefont {X.-G.}\ \bibnamefont {{Wen}}},\ }\href {\doibase
  10.1103/PhysRevB.82.165119} {\bibfield  {journal} {\bibinfo  {journal}
  {\prb}\ }\textbf {\bibinfo {volume} {82}},\ \bibinfo {eid} {165119} (\bibinfo
  {year} {2010})},\ \Eprint {http://arxiv.org/abs/1003.1774} {arXiv:1003.1774
  [cond-mat.str-el]} \BibitemShut {NoStop}%
\bibitem [{\citenamefont {{Swingle}}\ and\ \citenamefont
  {{Wen}}(2010)}]{2010arXiv1001.4517S}%
  \BibitemOpen
  \bibfield  {author} {\bibinfo {author} {\bibfnamefont {B.}~\bibnamefont
  {{Swingle}}}\ and\ \bibinfo {author} {\bibfnamefont {X.-G.}\ \bibnamefont
  {{Wen}}},\ }\href@noop {} {\bibfield  {journal} {\bibinfo  {journal} {ArXiv
  e-prints}\ } (\bibinfo {year} {2010})},\ \Eprint
  {http://arxiv.org/abs/1001.4517} {arXiv:1001.4517 [cond-mat.str-el]}
  \BibitemShut {NoStop}%
\bibitem [{Note2()}]{Note2}%
  \BibitemOpen
  \bibinfo {note} {A related concept for a finite group $G$, is $G$-injective
  PEPS \cite {SchuchEtal}. A $G$-injective tensor is a tensor which is
  invariant under a $G$-action on all inner indices simultaneously, together
  with the property that one can achieve any action on the virtual indices by
  acting on the physical indices. It was shown that these tensors are ground
  states of a parent Hamiltonian and have the topological entanglement entropy
  $\gamma = \protect \qopname \relax o{log}|G|$. This class of PEPS describe
  the universality class of quantum double models $D(G)$. Recently this
  concepts was generalized to $(G,\omega )$-injective PEPS \cite
  {Buerschaper:2013nga}, where the action of $G$ on the tensors are twisted by
  a 3-cocycle of $\omega $ of $G$. It was shown that these PEPS describes
  topological order in the university class of Dijkgraaf-Witten TQFT's \cite
  {Dijkgraaf:1989pz} and only depend on the cohomology class $[\omega ]\in
  H^3(G,U(1))$ of $\omega $.}\BibitemShut {Stop}%
\bibitem [{\citenamefont {{Schuch}}\ \emph {et~al.}(2010)\citenamefont
  {{Schuch}}, \citenamefont {{Cirac}},\ and\ \citenamefont
  {{P{\'e}rez-Garc{\'{\i}}a}}}]{SchuchEtal}%
  \BibitemOpen
  \bibfield  {author} {\bibinfo {author} {\bibfnamefont {N.}~\bibnamefont
  {{Schuch}}}, \bibinfo {author} {\bibfnamefont {I.}~\bibnamefont {{Cirac}}}, \
  and\ \bibinfo {author} {\bibfnamefont {D.}~\bibnamefont
  {{P{\'e}rez-Garc{\'{\i}}a}}},\ }\href {\doibase 10.1016/j.aop.2010.05.008}
  {\bibfield  {journal} {\bibinfo  {journal} {Annals of Physics}\ }\textbf
  {\bibinfo {volume} {325}},\ \bibinfo {pages} {2153} (\bibinfo {year}
  {2010})},\ \Eprint {http://arxiv.org/abs/1001.3807} {arXiv:1001.3807
  [quant-ph]} \BibitemShut {NoStop}%
\bibitem [{\citenamefont {Buerschaper}(2013)}]{Buerschaper:2013nga}%
  \BibitemOpen
  \bibfield  {author} {\bibinfo {author} {\bibfnamefont {O.}~\bibnamefont
  {Buerschaper}},\ }\href@noop {} {\  (\bibinfo {year} {2013})},\ \Eprint
  {http://arxiv.org/abs/1307.7763} {arXiv:1307.7763 [cond-mat.str-el]}
  \BibitemShut {NoStop}%
%%CITATION = ARXIV:1307.7763;%%
\bibitem [{\citenamefont {{He}}\ \emph {et~al.}(2014)\citenamefont {{He}},
  \citenamefont {{Moradi}},\ and\ \citenamefont {{Wen}}}]{HeMoradiWen}%
  \BibitemOpen
  \bibfield  {author} {\bibinfo {author} {\bibfnamefont {H.}~\bibnamefont
  {{He}}}, \bibinfo {author} {\bibfnamefont {H.}~\bibnamefont {{Moradi}}}, \
  and\ \bibinfo {author} {\bibfnamefont {X.-G.}\ \bibnamefont {{Wen}}},\
  }\href@noop {} {\bibfield  {journal} {\bibinfo  {journal} {ArXiv e-prints}\ }
  (\bibinfo {year} {2014})},\ \Eprint {http://arxiv.org/abs/1401.5557}
  {arXiv:1401.5557 [cond-mat.str-el]} \BibitemShut {NoStop}%
\bibitem [{Note3()}]{Note3}%
  \BibitemOpen
  \bibinfo {note} {In general we do not have a gauge theory and need another
  way to find the modular matrices in the right basis. One way is to find the
  basis which diagonalize $T$ and the $S$ matrix satisfy the requirements
  \begin {align*} &1.\protect \tmspace +\thickmuskip {.2777em}\protect \tmspace
  +\thickmuskip {.2777em} \protect \mathcal N^c_{ab} = \DOTSB \sum@ \slimits@
  _x\protect \frac {S_{ax}S_{bx}S^*_{cx}}{S_{1x}}\protect \tmspace
  +\thickmuskip {.2777em}\protect \text { is a positive integer},\\ &2.\protect
  \tmspace +\thickmuskip {.2777em}\protect \tmspace +\thickmuskip {.2777em}
  S_{ab} = S_{ba},\\ &3.\protect \tmspace +\thickmuskip {.2777em}\protect
  \tmspace +\thickmuskip {.2777em} S_{1,a}>0. \end {align*} In \cite
  {2013arXiv1303.0829L} it was shown for several examples, that this basis is
  unique and lead to the right form of $S$ and $T$.}\BibitemShut {Stop}%
\bibitem [{Note4()}]{Note4}%
  \BibitemOpen
  \bibinfo {note} {See \cite {SchulzetAlZn} for analysis of models of this
  type.}\BibitemShut {Stop}%
\bibitem [{\citenamefont {Trott}(1962)}]{trott1962pair}%
  \BibitemOpen
  \bibfield  {author} {\bibinfo {author} {\bibfnamefont {S.~M.}\ \bibnamefont
  {Trott}},\ }\href@noop {} {\bibfield  {journal} {\bibinfo  {journal}
  {Canadian Mathematical Bulletin}\ }\textbf {\bibinfo {volume} {5}},\ \bibinfo
  {pages} {245} (\bibinfo {year} {1962})}\BibitemShut {NoStop}%
\bibitem [{\citenamefont {Moradi}\ and\ \citenamefont {Wen}(2014)}]{MoradiWen}%
  \BibitemOpen
  \bibfield  {author} {\bibinfo {author} {\bibfnamefont {H.}~\bibnamefont
  {Moradi}}\ and\ \bibinfo {author} {\bibfnamefont {X.-G.}\ \bibnamefont
  {Wen}},\ }\href@noop {} {\bibfield  {journal} {\bibinfo  {journal} {to
  Appear}\ } (\bibinfo {year} {2014})}\BibitemShut {NoStop}%
\bibitem [{\citenamefont {Callan}\ and\ \citenamefont
  {Harvey}(1985)}]{Callan:1984sa}%
  \BibitemOpen
  \bibfield  {author} {\bibinfo {author} {\bibfnamefont {J.}~\bibnamefont
  {Callan}, \bibfnamefont {Curtis~G.}}\ and\ \bibinfo {author} {\bibfnamefont
  {J.~A.}\ \bibnamefont {Harvey}},\ }\href {\doibase
  10.1016/0550-3213(85)90489-4} {\bibfield  {journal} {\bibinfo  {journal}
  {Nucl.Phys.}\ }\textbf {\bibinfo {volume} {B250}},\ \bibinfo {pages} {427}
  (\bibinfo {year} {1985})}\BibitemShut {NoStop}%
%%CITATION = NUPHA,B250,427;%%
\bibitem [{\citenamefont {Levin}(2013)}]{L1309}%
  \BibitemOpen
  \bibfield  {author} {\bibinfo {author} {\bibfnamefont {M.}~\bibnamefont
  {Levin}},\ }\href {\doibase 10.1103/PhysRevX.3.021009} {\bibfield  {journal}
  {\bibinfo  {journal} {Phys. Rev. X}\ }\textbf {\bibinfo {volume} {3}},\
  \bibinfo {pages} {021009} (\bibinfo {year} {2013})},\ \Eprint
  {http://arxiv.org/abs/arXiv:1301.7355} {arXiv:1301.7355} \BibitemShut
  {NoStop}%
\bibitem [{\citenamefont {Dijkgraaf}\ and\ \citenamefont
  {Witten}(1990)}]{Dijkgraaf:1989pz}%
  \BibitemOpen
  \bibfield  {author} {\bibinfo {author} {\bibfnamefont {R.}~\bibnamefont
  {Dijkgraaf}}\ and\ \bibinfo {author} {\bibfnamefont {E.}~\bibnamefont
  {Witten}},\ }\href {\doibase 10.1007/BF02096988} {\bibfield  {journal}
  {\bibinfo  {journal} {Commun.Math.Phys.}\ }\textbf {\bibinfo {volume}
  {129}},\ \bibinfo {pages} {393} (\bibinfo {year} {1990})}\BibitemShut
  {NoStop}%
%%CITATION = CMPHA,129,393;%%
\bibitem [{\citenamefont {{Liu}}\ \emph {et~al.}(2013)\citenamefont {{Liu}},
  \citenamefont {{Wang}}, \citenamefont {{You}},\ and\ \citenamefont
  {{Wen}}}]{2013arXiv1303.0829L}%
  \BibitemOpen
  \bibfield  {author} {\bibinfo {author} {\bibfnamefont {F.}~\bibnamefont
  {{Liu}}}, \bibinfo {author} {\bibfnamefont {Z.}~\bibnamefont {{Wang}}},
  \bibinfo {author} {\bibfnamefont {Y.-Z.}\ \bibnamefont {{You}}}, \ and\
  \bibinfo {author} {\bibfnamefont {X.-G.}\ \bibnamefont {{Wen}}},\ }\href@noop
  {} {\bibfield  {journal} {\bibinfo  {journal} {ArXiv e-prints}\ } (\bibinfo
  {year} {2013})},\ \Eprint {http://arxiv.org/abs/1303.0829} {arXiv:1303.0829
  [cond-mat.str-el]} \BibitemShut {NoStop}%
\end{thebibliography}%
